# Anharmonic Self-Energy of Phonons: Ab Initio Calculations and Neutron Spin Echo Measurements

A. Debernardi,[1] F. de Geuser,[2] J. Kulda,[2] M. Cardona,[3] and E. E. Haller[4]

[1] Istituto Nazionale per la Fisica della Materia, Trieste University and SISSA research unit, Strada Costiera 12, Trieste, Italy
[2] Institut Laue-Langevin, B.P. 156, 38042 Grenoble Cedex 9, France
[3] Max-Planck-Institut für Festkörperforschung, Heisenbergstrasse 1, D-70569, Stuttgart, Germany
[4] Department of Materials Science and Engineering, Mail Code 176, 553 Evans Hall, University of California, Berkeley, CA 94720, USA

**Abstract.** We have calculated (*ab initio*) and measured (by spin-echo techniques) the anharmonic self-energy of phonons at the X-point of the Brillouin zone for isotopically pure germanium. The real part agrees with former, less accurate, high temperature data obtained by inelastic neutron scattering on natural germanium. For the imaginary part our results provide evidence that transverse acoustic phonons at the X-point are very long lived at low temperatures, i.e. their probability of decay approaches zero, as a consequence of an unusual decay mechanism allowed by energy conservation.

## 1. Introduction

The anharmonic self-energies (SE's) of phonons in semiconductors have been the object of considerable recent interest [1, 2, 3], stimulated in part by *ab initio* calculations [4, 5, 6, 7]. This SE determines two properties of collective excitations in the crystal which depend on the anharmonicity of the adiabatic interatomic potentials: the temperature dependence of the frequencies of phonons and their lifetimes. The latter provide one of the main microscopic mechanisms that govern the evolution of excited solids towards thermal equilibrium. As an effect of anharmonic interaction, a non-equilibrium population of vibrational modes (created either by the annihilation of hot carriers or by electromagnetic radiation) can decay into phonons of lower energy or can be scattered by a thermal phonon into phonons of different frequencies.

For these reasons the phonon SE's have been the subject of several experimental studies [8]. The poor resolution achieved with conventional inelastic neutron scattering (INS) has restricted such experimental work mainly to first-order Raman spectroscopy, to which only phonons close to the center of the Brillouin zone (BZ) are accessible. Recent developments in the spin-echo technique of inelastic neutron scattering [9, 10] have, however, opened the way to more accurate determination of the real and imaginary parts of the phonon SE's at other points in the BZ.

We report calculated (*ab initio*) and measured (by neutrons spin-echo techniques) SE's at the X-point of the BZ for natural and isotopically pure germanium. The calculated real part agrees with the spin-echo results as well as with former, less accurate, high temperature data obtained by classical INS [11]. Agreement between the calculated imaginary part and the neutron spin echo data is also good.

Our results suggest that in germanium a non-equilibrium transverse acoustic (TA) phonon population at the X-point is rather stable at low temperatures (i.e. it has a very small decay probability). Our study provides a complete analysis of the microscopic mechanisms involved in the decay process.

## 2. Phonon Self-Energy

Within the harmonic approximation, phonons are non-interacting and live forever [12]. Anharmonic effects manifest themselves as phonon-phonon interaction. The physical quantity that describes these effects is the phonon SE, $\Delta(\omega) - i\Gamma(\omega)$, which is a complex function [13]. For a phonon of frequency $\Omega$ the real part, $\Delta(\Omega)$, is associated with the change of the frequency (i.e. renormalization) due to scattering by other phonons, and thus is responsible for the temperature dependence of the phonon frequency. The probability of phonon decay, described by the imaginary part $\Gamma(\Omega)$, represents the inverse of the phonon lifetime. The latter quantity can be measured directly in time resolved experiments [14] while the linewidth can be measured by Raman spectroscopy [8, 15]. If the Raman peak displays a Lorentzian lineshape, basically the full width at half maximum (FWHM) of the phonon line, is equal to $2\Gamma$.

In the following we will use equivalently both the terms lifetime and linewidth to indicate the same physical property. The computation of phonon SE's can be performed in a density functional framework by means of perturbation theory according to the methods given in Refs. [6] and [7] where the interested reader can find technical details.

At equilibrium the atoms of the crystal are at the positions $\mathbf{R} + \tau_s$, where $\mathbf{R}$ denotes a direct lattice vector, while $s$ labels the different atoms of the unit cell. We find it useful to introduce the following notation in which the formulas obtained by perturbation theory can be written in a compact form by using second quantization phonon creation and destruction operators of the $j$-th mode $a_j^+(q)$ and $a_j(q)$. The operator corresponding to the atomic displacements from equilibrium position reads

$$[\mathbf{u}_{s,\alpha}(\mathbf{R})]_j = \sum_q \left(\frac{\hbar}{2\omega_j(\mathbf{q})M_s N}\right)^{\frac{1}{2}} [\mathbf{e}(\mathbf{q};j)]_{s,\alpha}\, e^{i\mathbf{q}\cdot\mathbf{R}} \left(a_j^+(-\mathbf{q}) + a_j(\mathbf{q})\right), \quad (1)$$

where $M_s$ is the mass of the $s$-th atom of the unit cell, $N$ the number of unit cells in the crystal, $\omega$ the frequency of a phonon of the $j$-th branch with wave vector $\mathbf{q}$ in the first Brillouin zone and $\mathbf{e}$ the eigenvector giving the amplitude of the corresponding mode. $[\mathbf{a}]_i$ denotes the $i$-th component of a vector $\mathbf{a}$. We define the "vector derivative" of a function $f$ as

$$\frac{\partial f}{\partial \mathbf{a}} \equiv \sum_i \frac{\partial f}{\partial [\mathbf{a}]_i} \cdot [\mathbf{a}]_i. \quad (2)$$

It is straightforward to generalize this definition to deal with higher order derivatives.

To the lowest order in the perturbation expansion of the total energy $E_\text{tot}$ with respect to the phonon amplitude, the width of a phonon of frequency $\omega$ and wave-vector $\mathbf{q}$ in the $j$-th branch can be written as:

$$\Gamma_j(\mathbf{q};\omega) = \frac{\pi}{2\hbar^2} \sum_{\mathbf{q_1},\mathbf{q_2},j_1,j_2} \left|\frac{\partial^3 E_\text{tot}}{\partial \mathbf{e}(\mathbf{q};j)\partial \mathbf{e}(\mathbf{q_1};j_1)\partial \mathbf{e}(\mathbf{q_2};j_2)}\right|^2 \boldsymbol{\Delta}(-\mathbf{q}+\mathbf{q_1}+\mathbf{q_2})\times$$
$$\{[n(\mathbf{q_1};j_1) + n(\mathbf{q_2};j_2) + 1]\,\delta(\omega - \omega_{j_1}(\mathbf{q_1}) - \omega_{j_2}(\mathbf{q_2})) +$$
$$2\,[n(\mathbf{q_1};j_1) - n(\mathbf{q_2};j_2)]\,\delta(\omega + \omega_{j_1}(\mathbf{q_1}) - \omega_{j_2}(\mathbf{q_2}))\}, \quad (3)$$

where $n(\mathbf{q}; j)$ is the Bose-Einstein thermal occupation factor of the $j$-th phonon mode with wave-vector $\mathbf{q}$. The function $\mathbf{\Delta}(\mathbf{q})$ is unity if the vector $\mathbf{q}$ is a reciprocal lattice vector and vanishes otherwise.

The first term in curly brackets on the right hand side (rhs) of Eq. 3 is responsible for the decay into two phonons of lower energy and we will refer to this mechanism as the *down-conversion* or *sum process*; the second term describes the process in which a non-equilibrium phonon is destroyed together with a thermal phonon and a phonon (of higher energy with respect to the initial phonons) is created, it will be called the *up-conversion* [14] or *difference process*.

The real part of the SE is composed of three contributions

$$\Delta_j(\omega) = \Delta_j^{(0)} + \Delta_j^{(3)}(\omega) + \Delta_j^{(4)}, \tag{4}$$

where, to simplify the notation, we have omitted the dependence on wave-vector (in the following we will use this convention whenever there is no possibility of confusion). Note that strictly speaking the self energy is only the term $\Delta_j^{(3)}(\omega)$ but we include in Eq. 4 the additional real contributions $\Delta_j^{(0)}$ and $\Delta_j^{(4)}$. The first term on the rhs of Eq. 4 represents the change in frequency due to the thermal expansion of the lattice [16]:

$$\Delta_j^{(0)} = \omega_j \left\{ \exp\left[-3\gamma_j \int_0^T \alpha(T')dT'\right] - 1 \right\} \tag{5}$$

where $\gamma_j$ is the $j$-mode Grüneisen parameter and $\alpha(T)$ the linear thermal expansion coefficient at temperature $T$. The second term in the rhs of Eq. 4 can be obtained through the Hilbert transform of $\Gamma_j(\mathbf{q}; \omega)$,

$$\Delta_j^{(3)}(\omega) = -\frac{2}{\pi} P \int_0^\infty \frac{\omega' \Gamma_j(\omega')}{(\omega'^2 - \omega^2)} d\omega' \tag{6}$$

where $P$ represents the Cauchy principal part of the integral. In the following we will refer to $\Delta_j^{(3)}(\omega)$ as the third order term.

The last term on the rhs of Eq. 4, which is real, is related to the fourth order derivative of the total energy through the expression:

$$\Delta_j^{(4)} = \frac{12}{\hbar} \sum_{\mathbf{q_1}, j_1} \frac{\partial^4 E_{\text{tot}}}{\partial \mathbf{e}(\mathbf{q}; j) \partial \mathbf{e}(-\mathbf{q}; j) \partial \mathbf{e}(\mathbf{q_1}; j_1) \partial \mathbf{e}(-\mathbf{q_1}; j_1)} [2n(\mathbf{q_1}; j_1) + 1] \tag{7}$$

The renormalized phonon frequency $\Omega_j$ that includes the anharmonic contributions is the solution of the Dyson equation:

$$\Omega_j^2 = \omega_j^2 + 2\omega_j \Delta_j(\Omega_j) \tag{8}$$

Note that in the usual case of small $\Delta_j(\Omega_j)$ Eq. 8 can be simplified to

$$\Omega_j = \omega_j + \Delta(\omega_j) \tag{9}$$

## 3. Computational details

We have computed the real and imaginary parts of the SE, $\Delta$ and $\Gamma$ by means of the density functional theory and of pseudopotential techniques [17]. We use a plane wave expansion of the electronic density with a kinetic energy cutoff of 20 Ry. Germanium has the crystal structure of diamond (FCC with two atoms per unit cell), however, in our calculations we have found it convenient to use a cell with a reduced symmetry containing four Ge atoms, in this way the phonon modes at the X-point of the FCC cell are folded back into the modes at the zone center of the double cell. The linear thermal expansion coefficient and the Grüneisen parameter appearing in Eq. 5 are computed in the same way as in Ref. [18]. The third order term is calculated using density functional perturbation theory [4, 6]. Technical details can be found in Ref. [6] and references therein. We obtain the fourth order term by combining frozen phonon calculations [7] and perturbation theory [19]. The harmonic phonon frequencies are obtained from density functional perturbation theory [20].

The integration in reciprocal space to compute electronic densities and related data is performed by special points techniques with a 888 Monkhorst-Pack [21] grid in the FFC first Brillouin zone that reduces to 12 special points [22] in the irreducible wedge of the tetragonal cell used in our calculations. The integration over surface of constant energy, imposed by the $\delta$ functions appearing in Eq. 3, is carried out by the tetrahedra technique [23].

## 4. Experimental technique

Our experiments have been performed using the spin-echo option TASSE [9] on the IN20 polarized neutron three-axis spectrometer at the Institut Laue-Langevin in Grenoble (France). The sample was a germanium single crystal, having a volume of about 5 cm$^3$ and grown by the Czochralski method from material enriched to 96.8% of the $^{74}$Ge isotope. This high isotopic purity eliminates phonon scattering due to mass disorder, which is one of the mechanisms reducing the phonon life-time. The spin-echo measurement technique, described in detail elsewhere [10], consists in employing the Larmor precession to provide a precise measure of small changes in kinetic energy of each neutron upon scattering in the sample. The variation of the mean energy transfer is reflected by the shift of the spin-echo phase,

$$\Delta\varphi = \tau_F \frac{\Delta E}{\hbar}, \quad (10)$$

where $\tau_F$ is the Fourier time, a parameter characterizing the sensitivity of the instrument. Its value is proportional to the field integral in the precession coils and to the cube of the neutron wavelength. Similarly, a finite linewidth of an excitation line introduces an additional energy spread into the ensemble of neutrons, resulting in an enhanced precession phase spread and hence in stronger damping of the echo amplitude.

In the present case, the spectrometer was operated with a fixed final neutron wave-number $k_f = 4.1$ Å$^{-1}$. The [0, 0, 0.8] transverse acoustic phonon mode to be investigated has been localized in the [3, 3, 1] Brillouin zone by conventional $Q = const.$ scans with precession fields off. After doing this, the spectrometer has been kept at a fixed configuration corresponding to the maximum of the phonon peak at $Q = [3, 3, 1.8]$ with an energy transfer of $\Delta E = 9.8$ meV and scans of the precession fields have been performed for currents corresponding to diverse Fourier times.

The widths presented in the right-hand panel of Fig. 1 correspond to the damping of the echo signal at Fourier times in the range between 0.012 and 0.04 ns. A constant value of 6.5 GHz has been subtracted from the observed values at all temperatures - this component includes both the instrumental resolution and the contribution of phonon damping due to scattering by crystal defects and the remaining mass disorder, all of them assumed to be independent of temperature.

## 5. Results

In Fig. 1 we report our experimental and calculated results for the temperature dependence of the real and imaginary parts of the SE of the TA-phonon mode in germanium at the X-point. The experimental data (denoted by diamonds) refer to the $(0, 0, 0.8)\frac{2\pi}{a_L}$ point of the conventional cell of side $a_L$, while the *ab initio* computation was performed at the X-point $(0,0,1)\frac{2\pi}{a_L}$ in order to avoid large super-cell calculations.

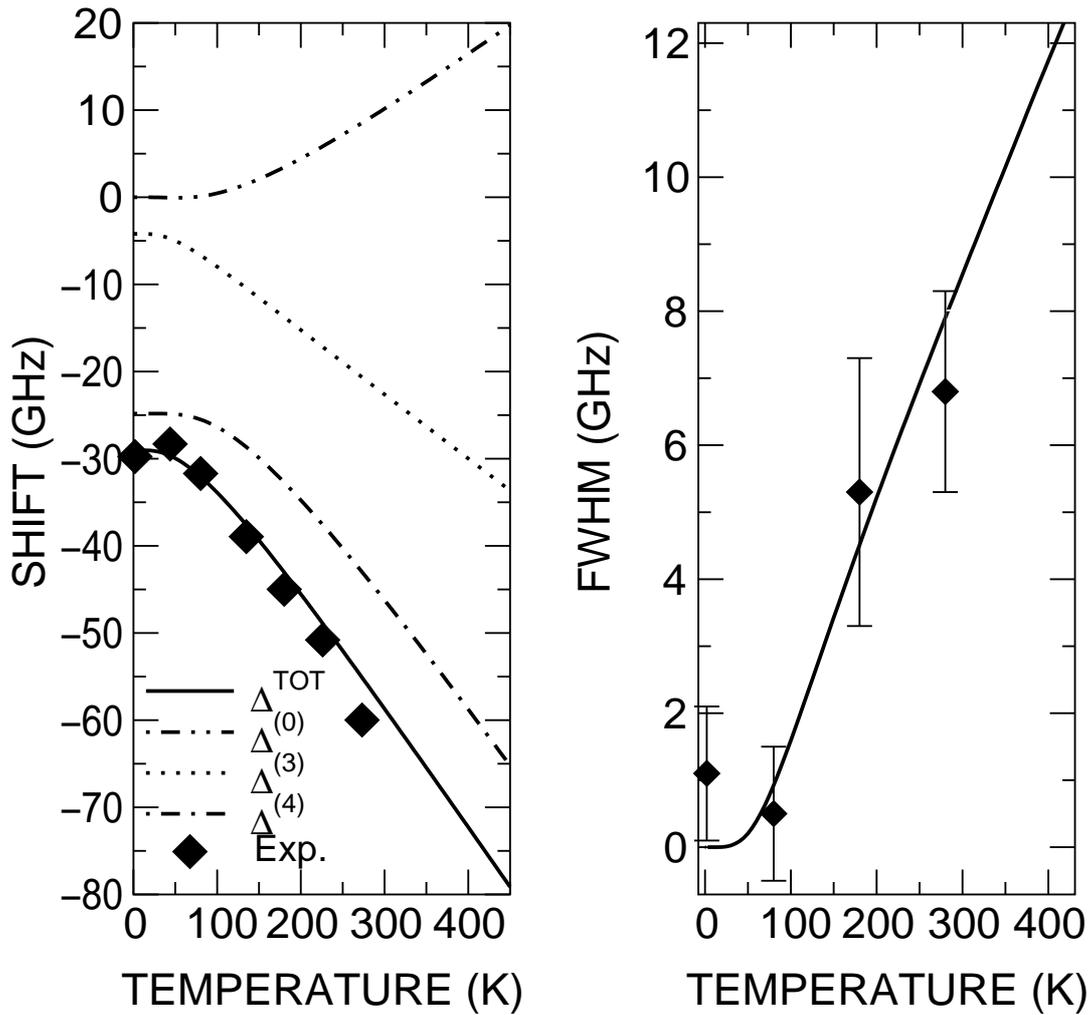

**Figure 1.** Real (left panel) and imaginary part (right panel) of the self-energy of the TA-phonon at X as a function of temperature. Diamonds denote spin-echo measurements; dashed and solid line denote *ab initio* calculation for real and imaginary part of the self-energy; the computed low order contributions to the real part of the self-energy are also displayed (see text) as dotted or dot-dashed lines. Note: $1 GHz \simeq 3.3 \times 10^{-2}$ cm$^{-1}$ $\simeq 41\mu$eV.

In the left panel of Fig. 1 we display our results for the real part of the phonon SE, i.e. the shift of the phonon frequency as a function of temperature. The results of *ab initio* calculation for the low order anharmonic contribution are also displayed and summed up ($\Delta^{\text{TOT}}$, dashed line) for comparison with experimental data (denoted by diamonds). As reported in the figure, the thermal expansion gives a positive contribution to the shift (denoted by $\Delta^{(0)}$) that added to the negative contribution of third and fourth order results in a small temperature dependence of the phonon frequency of the TA-mode under consideration. We note that at zero temperature $\Delta^{(3)}$ is much smaller than $\Delta^{(4)}$, just the opposite of what is found for the $\Gamma$-modes in most of the III-V semiconductors [6].

In the right panel we display the decay probability (linewidth), i.e. twice the imaginary part of the SE. Our first principles calculation agrees reasonably well with the spin-echo data.

The first important result is that the decay probability vanishes at zero temperature:

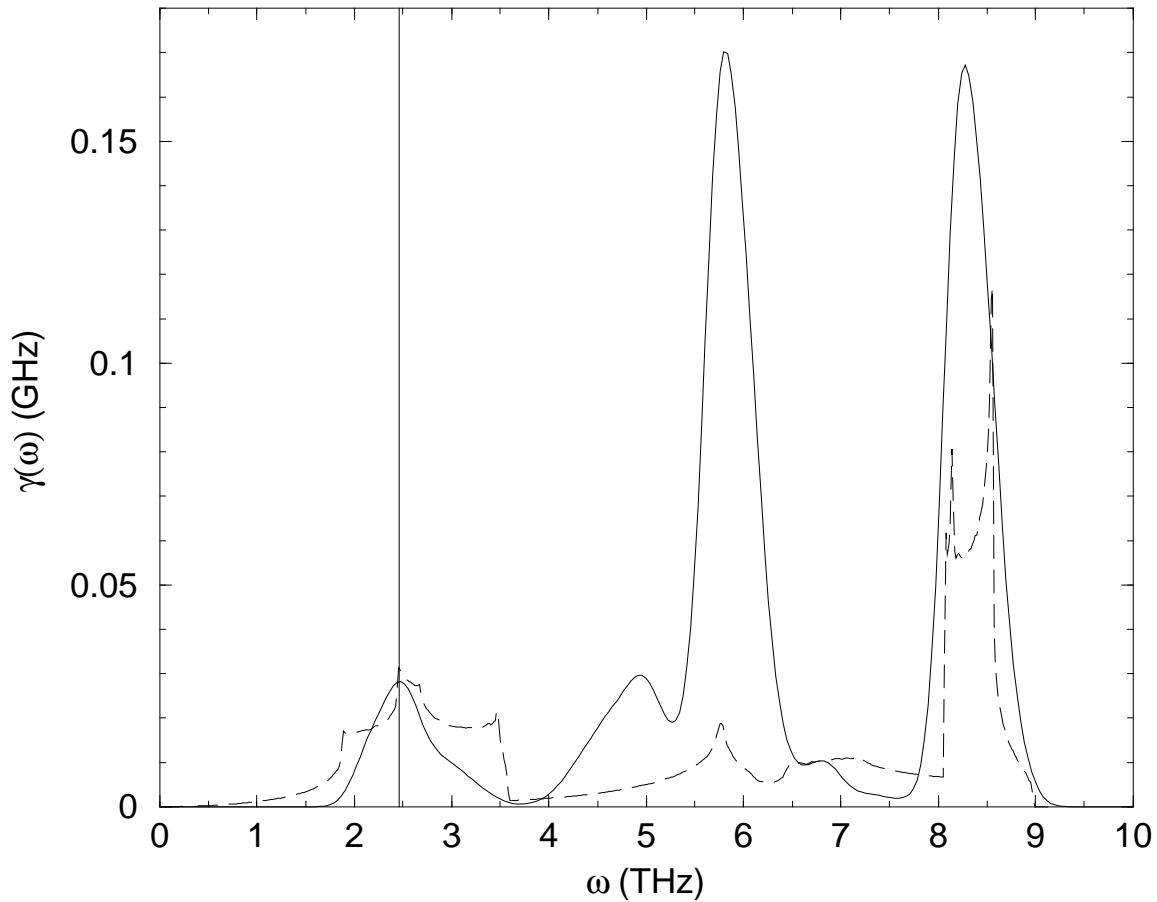

**Figure 2.** Solid line: frequency resolved final state spectrum $\gamma(\omega)$ calculated for $T = 300$ K. $\omega$ is the frequency of one of the two phonons involved in the difference decay process of the TA(X)-phonons [25]. Notice that the difference in the frequencies of the two main peaks equals the TA(X) frequency. The same happens for the two other peaks at $\sim 2.5$ and $5$ THz. Dotted line: one-phonon density of states. Vertical line: transverse acoustic phonon frequency.

the phonon is long-lived and the lifetime is not related to anharmonic components of the potential energy (the effect of decay into two phonons amounts to less than 0.1 GHz as shown in Fig. 1) but to other mechanisms, such as crystal defects or impurities. The vanishing of the decay probability, when the temperature tends to zero, follows from the absence of thermal phonon population at low temperatures. Here the only microscopic process allowed by energy conservation is the decay into acoustic phonons of lower energy whose wave vector is close to the Brillouin zone center (i.e. in the region where the two phonon branches depend linearly on the wave vector). Not only the density of final states is small for such processes but, as a consequence of translational symmetry, the corresponding matrix elements must vanish when the wave vector of the created phonon tends to zero [24]. According to our calculations, the down-conversion contribution remains negligible in all the range of temperatures studied. The only significant contributions arise from the up-conversion processes that involve the destruction of one thermal phonon; this also explains the narrow linewidths (less than the experimental resolution) measured by thermal neutron spectroscopy by Nelin and Nilsson [11].

To further analyze our results we have defined the *frequency resolved final state spectrum*, $\gamma(\omega)$ giving the probability per unit time that the TA phonon decays into one mode of frequency $\omega$ while the frequency of the other mode involved is determined by the energy conservation law, given by the delta functions in Eq. 3.

In practice, $\gamma(\omega)$ is obtained by restricting the sum over **q** and $j$ in Eq. 3 to those values for which $\omega_j(\mathbf{q}) = \omega$, by inserting $\delta(\omega - \omega_j(\mathbf{q}))$ under the summation sign. By definition the integral of $\gamma(\omega)$ over the whole range of frequencies gives the decay probability $\Gamma$.

In Fig. 2 we display $\gamma(\omega)$ calculated for the TA (X)-phonon at $T = 300$ K. The down-conversion contribution, giving peaks placed symmetrically with respect to the frequency $\omega_{\text{TA}}(X)/2$, is practically absent for the reasons discussed above. The up-conversion process, instead, gives a couple of peaks of equal height spaced by $\omega_{\text{TA}}(X)$. The lower energy peak of a given pair corresponds to the phonon destroyed in the decay process while the higher one to the created phonon. The situation here is opposite to what is found for the optical mode at the Brillouin zone center where only the down conversion mechanism is significant [4, 6].

In Fig. 2 we have also displayed the computed one phonon density of state (PDOS). The most important decay mechanism involves the destruction of a longitudinal phonon at the zone boundary (note that a van Hove singularity in the PDOS corresponding to these modes has the same $\omega$ of one of the two main peaks of $\gamma(\omega)$). The vertical line in Fig. 2 corresponds to the frequency of the TA-mode at X, and coincides with a singularity in the PDOS. This degeneracy is responsible for the two small peaks (roughly at 2.5 and 5.0 THz) and suggests the possibility of an unusual decay process for the TA(X)-phonon. This decay mechanism is quite sensitive to small contributions (such as higher order corrections) neglected in the present calculation and can affect the accuracy of our theoretical results at high temperatures.

## 6. Conclusion

We have compared first principles calculations and spin-echo measurements of the complex SE of the TA(X)-phonon of germanium. From our analysis of microscopic mechanisms we have found that the decay of this phonon is influenced by up-conversion, i.e. difference processes that result into a very long lived mode at low temperature.

## 7. Acknowledgment

We thank the Istituto Nazionale per la Fisica della Materia for the "Iniziativa Trasversale di Calcolo Parallelo", and appreciate the invaluable help of E. Farhi (ILL) with setting up the TASSE option on IN20.